\documentclass{aa}
\usepackage{graphicx}
\input{psfig}
\begin{document}

\title{Asteroseismology and calibration of $\alpha$\,Cen binary system}
\author{F. Th\'evenin \inst{1},
J. Provost \inst{1},
P. Morel \inst{1},
G. Berthomieu \inst{1},
F. Bouchy \inst{2},
\and
F. Carrier \inst{2}}
\institute{
D\'epartement Cassini, UMR CNRS 6529, Observatoire de la C\^ote 
d'Azur, BP 4229, 06304 Nice CEDEX 4, France.
\and
Observatoire de Gen\`eve, 51, chemin des Maillettes, 1290 Sauverny, Switzerland}
\titlerunning{Asteroseismology of $\alpha$\,Cen}
\authorrunning{Th\'evenin et al.}

\offprints{F. Th\'evenin}
\mail{Frederic.Thevenin@obs-nice.fr}

\date{Received date / Accepted date}

\abstract{
Using the oscillation frequencies of $\alpha$\,Cen\,A recently 
discovered by Bouchy \& Carrier~(\cite{bc01}, \cite{bc02}),
the available astrometric, photometric and spectroscopic
data, we tried to improve the calibration of the
visual binary system $\alpha$\,Cen.
With the revisited masses of Pourbaix et al.~(2002)
we do not succeed to obtain a
solution satisfying all the seismic observational constraints.
Relaxing the constraints on the masses, we have found
an age $t_{\rm\alpha\,Cen}=4\,850\pm500$\,Myr,
an initial helium mass fraction $Y_{\rm i} = 0.300\pm0.008$,
and an initial metallicity $(Z/X)_{\rm i}=0.0459\pm0.0019$,
with $M_{\rm A}=1.100\pm0.006\,M_\odot$ and
$M_{\rm B}=0.907\pm0.006\,M_\odot$ for $\alpha$\,Cen\,A\,\&\,B.
\keywords{
Stars: binaries: visual - Stars: evolution - Stars: oscillation -
Stars: fundamental parameters - Stars: individual: $\alpha$\,Cen
}}

\maketitle

\section{Introduction}\label{sec:int}
Over the last decade, many 
efforts to derive accurate fundamental parameters of the
double star $\alpha$\,Cen\,A\,\&\,B (HD128620/1) and to predict asteroseismic
frequencies have been carried on (e.g. Guenther \& Demarque~\cite{gd00}; 
Morel et al.~\cite{mpl00} and references therein).
For $\alpha$\,Cen\,A, the first frequency measurements  done by
Bouchy \& Carrier~(\cite{bc01}) exhibit discrepancies with 
the past predicted frequencies of the calibrated 
system.
The comparison of published theoretical frequencies 
(Morel et al.~\cite{mpl00}) with those deduced from the observations suggests 
that the discrepancies come from the adopted value for
the mass of $\alpha$\,Cen\,A.
Recently, taking into account the gravitational red-shift and 
the convective blue-shift, Pourbaix et al.~(\cite{pnm02}) have revisited
their previous analysis of astrometric and spectroscopic data. As a result,
the masses of stars $\alpha$\,Cen\,A \& B deviate significantly 
from their previous determination by more than $2\sigma$; new mass values:
$M_{\rm A}=1.105\pm0.007\,M_\odot$ and $M_{\rm B}=0.934\pm0.006\,M_\odot$
are lower than their old ones.
In this letter we present attempts to calibrate the binary system
taking into account all these new observational constraints.

\section{New constraints on the $\alpha$\,Cen\,binary}
\subsection{Luminosities.} For both components we consider the effective
temperatures
 $T_{\rm eff}$ and metallicities derived in our previous work
(Morel et al.~\cite{mpl00}). 
We derive the luminosities from the accurate Geneva photometry
(Burki et al.~\cite{b02}). The magnitudes
 $V_{\rm A}=-0.003\pm 0.004$ and $V_{\rm B}=1.332\pm 0.005$ are combined with
the new value of the parallax $\varpi=747.1\pm1.2$\,mas
(Pourbaix et al.~\cite{pnm02})
to derive the luminosities $L_{\rm A}$ and $L_{\rm B}$. 
The bolometric corrections 
used are from Flower's~(\cite{fl96}). 
Lejeune et al.~(\cite{lcb98}) and Bessell et al.~(\cite{bcp98})
bolometric corrections have been tried 
in order to estimate the uncertainty on the derived luminosities. 
Both range the luminosity values derived from Flower's calibration
leading us to adopt these last one with an uncertainty of $0.018$ and
 $0.016$ (solar unit) respectively for $L_{\rm A}$ and $L_{\rm B}$ as reported 
in Table~\ref{tab:mod}.

\begin{figure}
\psfig{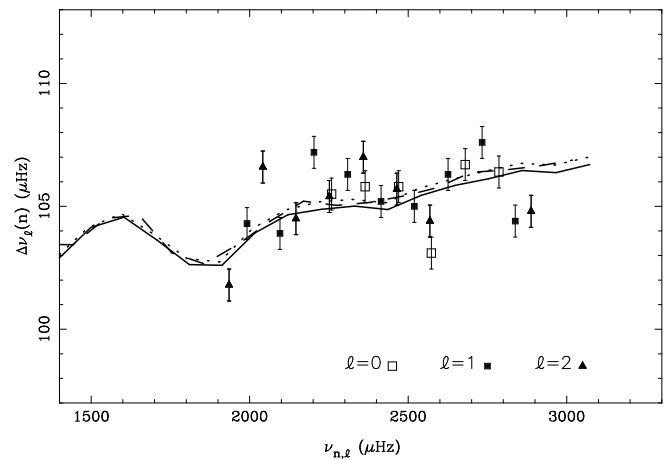}
\caption{
Large frequency spacing as a function of the frequency.
Symbols indicate the Bouchy \& Carrier (2002) observed values
with their error bars.
Continuous lines correspond to the model with $M_{\rm A}=1.100\,M_\odot$.
Full, dashed and dotted lines correspond respectively to modes $\ell$=0, 1
and 2.
}\label{fig:large}
\end{figure}

\begin{figure}
\psfig{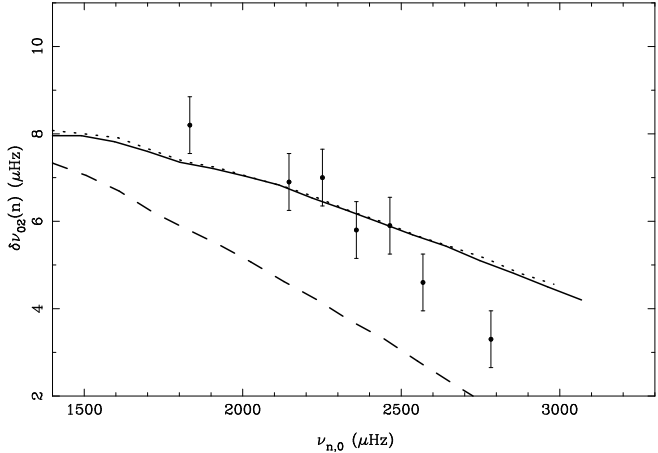}
\caption{
Small frequency spacing $\delta\nu_{0,2}$ as a function of the frequency.
Symbols indicate the observed values with their error bars.
Full, dashed and dotted lines correspond respectively to models with
$M_{\rm A}=1.100\,M_\odot$, $M_{\rm A}=1.105\,M_\odot$ (mass of Pourbaix et 
al. 2002) and $M_{\rm A}=1.114\,M_\odot$ (without convective core).
}\label{fig:sdeux}
\end{figure}

\begin{figure}
\psfig{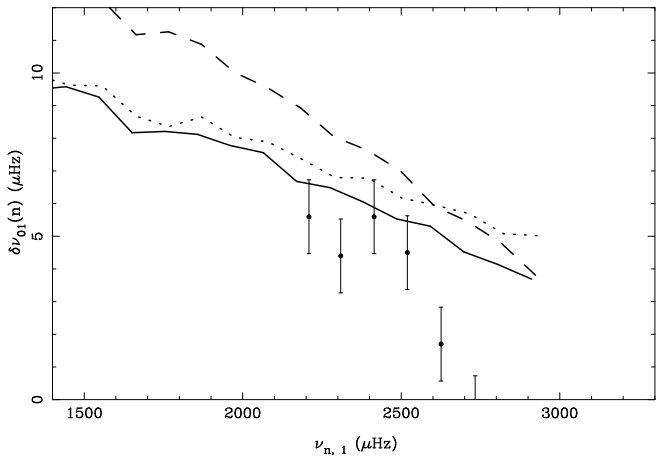}
\caption{
Same as Fig.~\ref{fig:sdeux} for
small frequency spacing $\delta\nu_{0,1}$.
}\label{fig:sun}
\end{figure}

\subsection{P-modes oscillations discovered.}\label{sec:osc} 
Recently solar-like $p$-mode oscillations in $\alpha$\,Cen\,A
have been detected by 
Bouchy \& Carrier~(\cite{bc01}) with {\sc coralie}
fiber-fed spectrograph. With a longer time serie of observations
Bouchy \& Carrier~(\cite{bc02}) have identified 28 oscillation frequencies
$\nu_{n,\ell}$ 
in the power spectrum of the velocity power spectrum, 
with degrees $\ell = 0,1,2$ and radial orders $n$ from 15 to 25.
The values of the frequencies depend strongly on the surface properties of
the star poorly described by the models. 
In seismological
analysis one rather uses to characterize the set of oscillation
frequency three frequency spacings,
one ``large'' and two ``small'', 
that are less surface dependent.
The large frequency spacing, difference between
frequencies of modes with consecutive radial order $n$:
$\Delta\nu_\ell(n) \equiv \nu_{n, \ell} - \nu_{n-1,\ell}.$ 
In the high frequency range, i.e. large radial orders, 
$\Delta\nu_\ell$ is almost constant with 
a mean value $\Delta_0$, strongly related to the mean density of the star.
The small separation, difference between frequencies of modes with 
degree of same parity and 
with consecutive radial order: $\delta\nu_{0,2}(n)\equiv\nu_{n,0}-\nu_{n-1,2}$ 
is very sensitive to the core of the star, i.e. to its age. 
Another small spacing
sensitive to the core is obtained by combining modes of degrees $\ell$=0 and 1:
$\delta\nu_{0,1}(n)\equiv\nu_{n+1,0}+\nu_{n,0}-2\nu_{n,1}.$
Figure~\ref{fig:large} shows the observed large spacing of $\alpha$\,Cen\,A
as a function of the frequency. 
Error bars of $\pm 0.65\,\mu$Hz have been determined considering the frequency 
resolution of the time series. The dispersion of the observed points according 
to the mode degree is larger than the error bars. As discussed in Bouchy \& 
Carrier (2002), it may be due partly to a possible systematic error of $\pm$ 1.3 
$\mu$Hz introduced at some identified mode frequencies, especially above 2.5 
mHz, by aliases and/or rotational splitting. 
Figures~\ref{fig:sdeux}~\&~\ref{fig:sun} show the small
spacings $\delta\nu_{0,2}$ and 
$\delta\nu_{0,1}$ as a function of the frequency.

\section{New evolutionary models}\label{sec:res}
This work is an extension of Morel et al.~(\cite{mpl00}),
taking into account the additional seismic constraints.
It consists in computing evolved models of  $\alpha$\,Cen\,A\,\&\,B
until they reach together at the same age,
the measured luminosity, effective 
temperature and metallicity. The
free parameters are the age $t_{\rm\alpha\,Cen}$,
the initial helium content $Y_{\rm i}$ and metallicity $(Z/X)_{\rm i}$ and
the mixing-length parameters $\lambda_{\rm A}$, $\lambda_{\rm B}$.
We assume that $\lambda_{\rm A} \equiv \lambda_{\rm B} \equiv\lambda$ because
both stars have similar masses and chemical abundances.
The relaxation of this constraint
 gives similar results  as emphasized in
the discussion of Morel et al.~(\cite{mpl00}).
All models have been computed with the {\sc cesam} 
code (Morel~\cite{m97}) -- see Morel et al.~(\cite{mpl00}) for details.
Models are initialized at the homogeneous {\sc zams}, using the 
Canuto \& Mazitelli~(\cite{cm91,cm92}) convection theory.

In a first step the new Pourbaix et al.~(\cite{pnm02}) masses are used as
observable constraints and without the seismic constraints.
The solution obtained by a $\chi^2$ fitting
gives $t_{\rm\alpha\,Cen}=8\,600$\,Myr, $Y_{\rm i}=0.256$,
$(Z/X)_{\rm i}=0.0459$ and $\lambda= 1.3$. 
This solution does not fulfill all available seismic constraints
for $\alpha$\,Cen\,A. The large spacing is well fitted by this solution
leading to estimates of the mean density and, at fixed mass,
to a radius $R_{\rm A}\approx 1.23\,R_\odot$. On the contrary, the
small frequency spacings (Fig.~\ref{fig:sdeux}~\&~\ref{fig:sun}) of
this model deviate significantly from the observations, leading us 
to reject this solution; $\delta\nu_{0,2}$
is too small, therefore the age is too large (Morel et al.~\cite{mpl00}).
Moreover, an age of $t_{\rm\alpha\,Cen}=8\,600$\,Myr is difficult to accept
for a star having a metallicity larger than solar.
Note also that the derived mixing length parameter, $\lambda=1.3$, is rather
large for
the convection theory of Canuto \& Mazitelli~(\cite{cm91}, \cite{cm92}) 
predicting values closer to unity.

\begin{table}
\caption[]{Characteristics
of $\alpha$\,Cen\,A \& B models.
The first four rows recall the observed and used effective temperatures in K,
metallicities, luminosities and mean large spacing $\Delta_0$
(in $\mu$\,Hz). Symbols are defined in text.
The five next rows present the deduced
calibration parameters and
the next ones show some characteristics of the model. 
At center,
$T_{\rm c}$, $\rho_{\rm c}$, $X_{\rm c}$, $Y_{\rm c}$ are respectively the
temperature (in M\,K), the density (in g\,cm$^{-3}$), the
hydrogen and the helium mass fractions. Indexes s, c, i, cz and co correspond
respectively to observed surface values, center values, initial values and 
convective envelope and core. R, T and $\rho$ are respectively
the radius, the temperature and the density.
}\label{tab:mod}
\begin{tabular}{lllll}  
\hline \\
                     &$\alpha$\,Cen\,A &$\alpha$\,Cen\,B \\ 
\\
%\\ \hline \\
$T_{\rm eff}$          &$5790\pm30$\,K   &$5260\pm50$\,K \\
$\rm [Fe/H]$     &$0.20\pm0.02$    &$0.23\pm0.03$ \\
$L/L_\odot$           &$1.519\pm0.018$  &$0.5002\pm0.016$\\
$\Delta_0$	       &$105.5\pm0.5$\\
\\ \hline \\
$t_{\rm\alpha\,Cen}$\,(Myr)  &\multicolumn{2}{c}{$4\,850\pm 500$} \\
$Y_{\rm i}$           &\multicolumn{2}{c}{$0.300\pm0.008$}\\
$(Z/X)_{\rm i}$  &\multicolumn{2}{c}{$0.0459\pm0.0019$}\\
${\lambda}$ 	      &\multicolumn{2}{c}{$0.98\pm0.04$}\\
$M/M_\odot$           &$1.100\pm0.006$  &$0.907\pm0.006$\\
\\ \hline \\
$R/R_\odot$	     &1.230   &0.857\\
$X_{\rm s}$	     &0.715   &0.694 \\

$Y_{\rm s}$	     &0.258  &0.277\\
$(Z/X)_{\rm s}$      &0.0384 &0.0417\\
$\rm[Fe/H]_s$        &0.195  &0.231\\
\\
$R_{\rm cz}/R_\star$ &0.725  &0.679\\
$T_{\rm cz}$	     &1.893  &2.802\\
$R_{\rm co}/R_\star$ &0.052 \\
\\
$T_{\rm c}$	     &19.00  &13.89 \\
$\rho_{\rm c}$       &177.1  &117.1 \\
$X_{\rm c}$	     &0.182  &0.428 \\
$Y_{\rm c}$	     &0.785  &0.539  \\
\\ \hline
\end{tabular}
\end{table}

\begin{figure*}
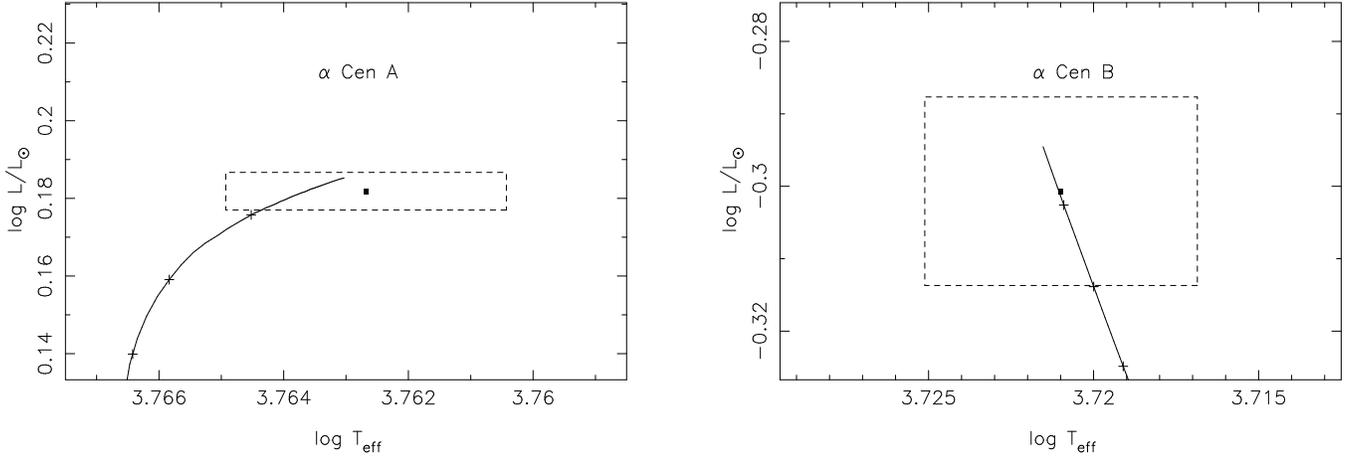

\vbox{\centerline{
\hbox{\psfig{figure=fig_thevenin4.ps,height=6.0cm,angle=270}
\hspace{1.truecm}
\psfig{figure=fig_thevenin5.ps,height=6.0cm,angle=270}}
}}
\caption{
Evolutionary tracks in the H-R diagram of models $\alpha$\,Cen\,A\,\&\,B.
Dashed rectangles delimit the uncertainty domains.
The ``+'' denote a path of 500\,Myr age.
}\label{fig:AB}
\end{figure*}

Thus, in a second step, we take into account the 
additional constraints given by observed small frequency spacings and we
consider the masses of the two stars as free parameters and no longer
as observational constraints. Our best solution comes up with
$t_{\rm\alpha\,Cen}=4\,850$\,Myr, $M_{\rm A}=1.100\,M_\odot$, 
$M_{\rm B}=0.907\,M_\odot$, $Y_{\rm i}=0.300$, 
$(Z/X)_{\rm i}=0.0459$ and $\lambda=0.98$. The sum of masses and
the fractional mass we derive are compatible with the astrometrical values of
Heintz~(\cite{h58}, \cite{h82}), Kamper \& Wesselink~(\cite{kw78}) and with the
values adopted in the calibration of Guenther \& Demarque~(\cite{gd00}).
Figure~\ref{fig:large}, \ref{fig:sdeux} and \ref{fig:sun} respectively
show the large and
small frequency spacings. We are aware that the two observed small spacings
above 2.5\,mHz
are considered as less reliable as discussed previously.
Table~\ref{tab:mod} gives characteristics of the corresponding
models of $\alpha$\,Cen\,A\,\&\,B. The confidence limits of
each calibration parameter,
the other being fixed, correspond to the maximum and minimum
values it can reach, 
in order that the generated models fit the observable targets within their 
error bars. Figure~\ref{fig:AB}
presents evolutionary tracks of two stars in the HR diagram.
Table~\ref{tab:freq} presents their $p$-mode frequencies in order to predict
large and small spacings for future observations.

At the age $t_{\rm\alpha\,Cen}=4\,850$\,Myr 
the model of $\alpha$\,Cen\,A presents a convective core
with still burning hydrogen. As emphasized by
Guenther \& Demarque~(\cite{gd00}) two kinds of model, with and without 
convective core, can satisfy the HR diagram constraints.
Indeed we have also found models of $\alpha$\,Cen\,A
without convective core satisfying the 
seismic constraints for $\Delta\nu_\ell$ and $\delta\nu_{0,2}$ but
they are ruled out by the $\delta\nu_{0,1}$ constraint.
As an example we have plotted in Fig.~\ref{fig:sdeux}~\&~\ref{fig:sun} the
small spacings for models $M_{\rm A}=1.114\,M_\odot$,
$M_{\rm B}=0.923\,M_\odot$,
corresponding to $t_{\rm\alpha\,Cen}=5\,170$\,Myr and $Y_{\rm i}$=0.285.
According to Guenther \& Demarque~(\cite{gd00}) 
models with and without convective core can  be 
discriminated by the so-called mode-bumped spacing;
our work shows that the small spacing $\delta\nu_{0,1}$ can be also
successfully used for this purpose.

\section{Discussion and conclusion}\label{sec:con}
Within the validity of the physics we use,
our classical calibration of the binary system $\alpha$\,Cen\,A\,\&\,B using
astrometric, photometric spectroscopic constraints with the dynamical masses of
Pourbaix et al.~(\cite{pnm02}) does not fully satisfy the seismic constraints
derived from the observations of Bouchy \& Carrier~(\cite{bc02}).
Relaxing the constraint on the masses we obtain a solution that agrees with
the seismic observations. The derived masses are close to those retained by 
Guenther \& Demarque~(\cite{gd00}). For $\alpha$\,Cen\,B
the difference, with respect to
Pourbaix et al.~(\cite{pnm02}) dynamical mass determination,
could indicate the presence of an unseen companion, a Jupiter
like planet or a brown dwarf, although its mass will be larger
than the upper limit given by Endl et al.~(\cite{ekehc01}). However,
it must be kept in mind that the masses we obtained are stellar model
dependent contrary to the astrometric masses which are mainly based
on the assumption of a purely Keplerian two-body problem.
Thus, the detection of oscillations of $\alpha$ Cen B 
are needed to better constrain its mass.
For this purpose Table~\ref{tab:freq} gives a set of expected
frequencies of this star
corresponding to a mean large spacing around $\Delta_0=162\,\mu$Hz. 
In addition, more accurate seismic observations of $\alpha$ Cen A 
are requested to decrease the dispersion of the large spacing
values and improve the small spacings.

Hopefully ground based experiments with {\sc coralie} and 
{\sc harps} fiber-fed
spectrograph (Bouchy \& Carrier~\cite{bc02b})
and the antarctic project {\sc concordiastro}
(Fossat et al.~\cite{fgv00}), and future space
missions like {\sc eddington} (Roxburgh~\cite{r02})
will provide accurate frequencies for
both components of our neighbour binary system.

\begin{table}
\caption[]{Low degree $p$-mode frequencies (in $\mu$\,Hz) 
for our calibrated models of $\alpha$\,Cen\,A\,\&\,B.}
\label{tab:freq}
\begin{tabular}{lllllllllllll}  
\hline \\
&\multicolumn{3}{c}{$\alpha$\,Cen\,A} & \multicolumn{3}{c}{$\alpha$\,Cen\,B}\\
\\ \hline \\
$n$& $\ell=0$&$\ell=1$&$\ell=2$&$\ell=0$&$\ell=1$&$\ell=2$ \\
 \\
    7&         860.3&         910.9&         962.5&
              1340.7&        1421.6&        1497.7\\
    8&         971.1&        1020.9&        1072.1&
              1514.7&        1595.5&        1671.5\\
    9&        1080.5&        1129.6&        1179.5&
              1688.2&        1767.1&        1841.8\\
   10&        1187.6&        1235.2&        1284.0&
              1858.1&        1936.2&        2008.9\\
   11&        1292.0&        1338.7&        1386.9&
              2024.4&        2101.1&        2173.6\\
   12&        1394.8&        1442.1&        1491.0&
              2188.6&        2264.7&        2337.0\\
   13&        1499.0&        1546.6&        1595.7&
              2351.7&        2428.7&        2501.2\\
   14&        1603.5&        1651.3&        1699.6&
              2515.4&        2592.3&        2665.1\\
   15&        1707.2&        1754.4&        1802.4&
              2678.8&        2755.1&        2827.2\\
   16&        1809.8&        1857.0&        1905.2&
              2840.3&        2916.6&        2988.4\\
   17&        1912.4&        1960.5&        2009.3&
              3001.0&        3076.9&        3149.3\\
   18&        2016.3&        2064.9&        2114.1&
              3161.6&        3237.9&        3310.6\\
   19&        2121.0&        2170.1&        2219.3&
              3322.5&        3399.6&        3472.8\\
   20&        2225.9&        2275.1&        2324.6&
              3484.4&        3561.4&        3635.1\\
   21&        2330.9&        2380.3&        2429.8&
              3646.4&        3723.6&        3797.4\\
   22&        2435.7&        2485.7&        2535.5&
              3808.2&        3885.8&        3960.0\\
   23&        2541.2&        2591.5&        2641.6&
              3970.6&        4048.3&        4124.0\\
   24&        2647.0&        2697.8&        2748.1&
              4133.2&        4211.4&        4286.4\\
   25&        2753.2&        2804.3&        2854.8&
              4296.4&        4374.9&        4450.4\\
   26&        2859.6&        2911.0&        2961.5&
              4460.1&        4538.7&        4614.5\\
   27&        2966.0&        3017.9&        3068.5&
              4624.0&        4703.0&        4779.0\\
   28&        3072.7&        3124.7&        3175.6&
              4788.2&        4867.4&        4943.9\\
\\ \hline
\end{tabular}
\end{table}

\begin{acknowledgements}
We would like to express our thanks to the referee Dr. Pourbaix, 
for helpful advices.  
This research has made use of the Simbad data base, operated at
CDS, Strasbourg, France and
of the WDS data base operating at USNO, Washington, DC USA.
It has been performed using the computing facilities provided by the OCA
program ``Simulations Interactives et Visualisation en Astronomie et
M\'ecanique (SIVAM)''. 
This work was in part financially supported by the Swiss National Science
Foundation.
\end{acknowledgements}

\stop
\end{document}